# Understanding the Effect of the COVID-19 Pandemic on the Usage of School Buildings in Greece Using an IoT Data-Driven Analysis

Georgios Mylonas
*Industrial Systems Institute*
*Athena Research & Innovation Center*
Patras, Greece
ORCID: 0000-0003-2128-720X

Dimitrios Amaxilatis
*SparkWorks ITC Ltd*
Derbyshire, United Kingdom
d.amaxilatis@sparkworks.net

Ioannis Chatzigiannakis
*Dpt. of Computer, Control & Management Engin.*
*Sapienza University of Rome*
Rome, Italy
ichatz@diag.uniroma1.it

*Abstract*—The COVID-19 pandemic has brought profound change in the daily lives of a large part of the global population during 2020 and 2021. Such changes were mirrored in aspects such as changes to the overall energy consumption, or long periods of sustained inactivity inside public buildings. At the same time, due to the large proliferation of IoT, sensors and smartphones in the past few years, we are able to monitor such changes to a certain degree over time. In this paper, we focus on the effect of the pandemic on school buildings and certain aspects in the operation of schools. Our study is based on data from a number of school buildings equipped with an IoT infrastructure. The buildings were situated in Greece, a country that faced an extended lockdown during both 2020 and 2021. Our results show that as regards power consumption there is room for energy efficiency improvements since there was significant power consumption during lockdowns, and that using other sensor data we can also infer interesting points regarding the buildings and activity during the lockdown.

## I. Introduction

The COVID-19 pandemic has proved to be one of the biggest challenges on a global scale since the Spanish flu of 1918 on many levels, including public health and the global economy. During 2020 and 2021, a very large part of the global population [1] experienced some form of lockdown or similar set of strict measures, at a scale believed to be unseen before. In this context, a large part of our daily activity has been affected to a variable degree, depending on the austerity of the lockdown and the spread of the pandemic. At the same time, this situation has provided an opportunity to the research community for monitoring directly the effect of these changes, or other phenomena that have surfaced as side effects of the pandemic and whose realization would otherwise a coordinated effort in order to achieve. An example of such aspects are energy and water consumption inside public and private buildings during the pandemic, or the pandemic's effect on air quality inside urban areas, just to name a few.

This has coincided with the fact that in the last few years, before the pandemic, we have seen the adoption of IoT and smart city technologies at a rapid pace throughout the world. The divide between the natural and the digital worlds has lessened considerably, and there are now multiple sensing endpoints in our offices, homes or even on our body, which continuously produce streams of sensor data. Privacy matters aside, this has made it tangibly easier to monitor the effects of the pandemic on a range of aspects of our everyday activity and the overall impact of the pandemic on our communities. As examples, Google [2] and Apple [3] produce reports for the changes in the mobility patterns of the users of mobile devices, based on requests to these companies' mapping services and mobile device location history data.

At the same time as the pandemic, there is an increasing interest in raising awareness about climate change and energy efficiency, with the EU announcing the European Green Deal [4] and a number of other countries making similar pledges. In this context, there has been a lot of interest on how the pandemic affected energy and water consumption overall, as well as in more specific areas such as public buildings, offices and homes. Regarding overall energy consumption, initial results that have shown a considerable drop in the first months of the pandemic, have contrasted with later results that showed a rebound in consumption [5].

In this setting of great interest about energy in general, the importance of the educational sector is self-evident, in terms of size and significance. Sustainable development and energy-saving behaviors are gradually becoming a part of educational programs. An example of recent activity to promote sustainability concepts in the educational sector through a structured curriculum is the United Nations' Climate Change Learning Partnership ( [6]). Simultaneously and in the context of the pandemic, as mentioned in [1], "schools, universities and colleges have closed either on a nationwide or local basis in 63 countries", a fact revealing a major disruption in the educational sector in terms of delivering classes, but also in terms of how schools function overall and consume resources, such as water and electricity.

In this work, we study the effect of the COVID-19 pandemic on certain aspects of the operation of a number of school buildings in Greece. We utilize the existing infrastructure and data produced by work kickstarted by the Green Awareness in Action [7] (GAIA) Horizon 2020 Project. GAIA produced a framework comprising IoT infrastructure in school buildings, applications, as well as educational content and action plans for increasing sustainability awareness among students, while also aiming for energy savings through behavioral changes in schools. Essentially, this infrastructure, which was deployed in most cases at least 1 year before the start of the pandemic, can provide us with data to help us draw comparisons regarding energy consumption, indoor comfort, indoor noise and air quality levels, among other.

In terms of research questions, and based on the data that we had available, we compiled the following list:

- How did the lockdown during 2020 and 2021 affect the power consumption inside this set of school buildings?
- Can use the GAIA dataset to detect potential behavior changes during these periods?
- Are there any other useful findings or correlations that we can infer using data like humidity, temperature and luminosity measurements?

Our results indicate that there is a considerable amount of power consumption inside the school buildings involved in our study, in the order of 20-40% of the normal power consumption during weekdays in most cases. As regards indoor noise level measurements, they can be used to track activity inside schools and correlate well with power consumption, but data at this point are inconclusive with respect to detecting behavior changes. With respect to other kind of measurements typically used in IoT deployments inside buildings, like temperature, humidity and luminosity, they can be used to infer some interesting points regarding the buildings themselves and activity during the lockdown.

## II. RELATED WORK

As mentioned in the previous section, although the pandemic is still ongoing, there is a lot of activity from the research community and, in relation to this work, specifically as regards energy consumption. An overview of the changes in energy consumption during 2020 at a global level was presented in [8]. Although there were some steep drops during the early part of the pandemic, energy consumption picked up pace during the second half of the year. At a global scale, a 6.4% overall reduction compared to 2019 was recorded, with areas like the EU and the US registering 7.7.% and 12.9% drops respectively. Similar results about the first lockdown period until April 2020 are also reported in [9], with a 17% reduction, with consumption returning to pre-COVID levels after mid-June 2020.

Regarding studies focusing on more specific examples of buildings, [10] presented a study on the energy use of several types of municipal buildings in Florianopolis, Brazil, during a period of almost 3.5 months of lockdown, discovering that almost half of the energy consumption in buildings like administrative buildings or elementary schools are not directly related to the presence of people inside them. Their findings to a certain degree agree with our own, as reported in the following sections. [11] focused on monitoring energy and hot water consumption patterns in a social housing building in Canada. The authors report that during the first, more strict, months of the lockdown overall consumption changed slightly, with the most notable change being the change in the time of the day as regards demand, noting that consumption moved to work hours instead of evening. A report [5] by the International Energy Agency (IEA) agrees with these findings, providing further details into the timing of the bounce of energy consumption back to 2019 levels in several countries

In this work, we focus on the effect of the pandemic on energy consumption specifically in school buildings, as well as thermal comfort and indoor noise levels. We focus on school buildings only in Greece, having identical or similar periods of lockdown and normal operation. Such aspects reflect either directly or indirectly the effect of the pandemic and lockdown on the behavior of school staff and students, as well as provide insights to the overall operation of the schools.

## III. THE GAIA PROJECT- INFRASTRUCTURE AND HARDWARE DESCRIPTION

The work presented here was conducted in the context of the GAIA Project [12]. Its main objective consisted in increasing awareness and promoting responsible behavior towards energy efficiency in educational communities. Sensors and data processing technologies were used to realize educational activities on top of data collected from an IoT infrastructure. Sensors were installed in school buildings to help students monitor the energy consumption of their building, or specific rooms, and become aware of the impact of environmental parameters and their behaviour as regards energy consumption. Overall, 25 schools (from primary to high school level) located in Greece, Italy and Sweden were involved.

This IoT infrastructure currently comprises over 1200 IoT monitoring endpoints and utilizes several hardware and software technologies [13], [14]. In each building, a set of sensors are deployed to monitor the following parameters: active power, energy, internal environmental parameters (e.g., luminosity, noise), external weather and pollution parameters. The actual deployment at each site is customized and adapted to the specific characteristics of the building (e.g., size, number of students, orientation, building plan). Measurements are continuously acquired by the GAIA IoT Platform which implements data processing services, with sensor data streams processed and aggregates extracted in real time.

## IV. METHODS

In this work, we focus on studying the effects of the pandemic on the school buildings in Greece for which we had available data both for the pandemic, as well as previous school years. This allows us to study such phenomena within a set of public schools that had similar changes in their operational environment, since they are located in the same country. Some differences in their operation were due to the different educational level of the schools, since we had data available from primary to high school level.

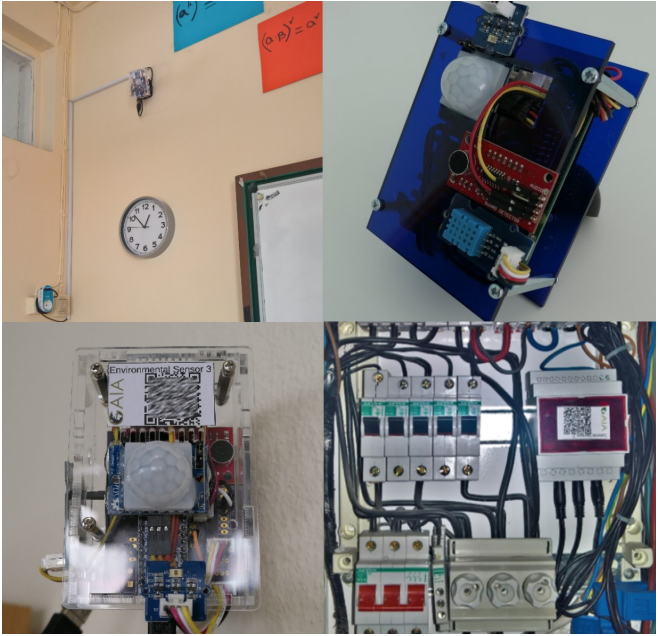

Fig. 1: Examples of nodes and installations in schools involved in the study.

The overall timeline of the pandemic, lockdown period and the respective effect on schools in Greece can be seen in Fig. 2. After some initial cases were reported, all schools were closed on March 10, 2020 and a strict lockdown began on March 23. Schools started to reopen gradually on May 11, until the end of the school year in June. After the summer break, schools reopened in September 14 and closed again during November, in 2 stages. On January 2021, schools reopened in 2 stages, while in some areas of the country schools were closed and reverted to remote class mode, since these areas were designated as "red" (i.e., with a high number of cases). On May 10, 2021, all schools in the country reopened, with any new closures decided for each class and school separately upon the detection of COVID-19 cases, based on a number of criteria related to the number of cases.

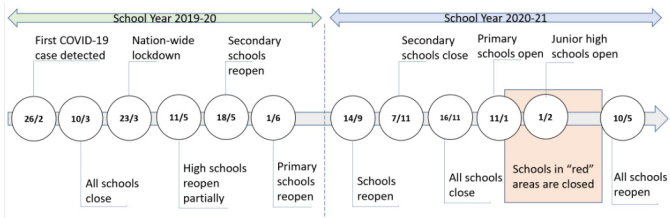

Fig. 2: Timeline of the pandemic in Greece and its effect on schools' operation from March 2020 to May 2021.

In terms of the question of building diversity and whether the school buildings we use in this study are representative of the schools in Greece, we should note that this set of buildings comprises schools in several parts of the country, including both urban and rural areas, as well as schools from mainland and island areas. They also include primary, junior high and high schools, covering a wide range of cases.

As mentioned in the previous sections, in terms of things that we will be investigating, these datasets first of all present an opportunity to study the baseline of schools in Greece with regards to power consumption. Although some conclusions regarding this aspect can be drawn by data from regular periods, the lockdown periods present an interesting scenario, since in several cases although lessons were entirely remotely, the buildings in many cases were open to use from the educational staff, especially in the looser periods of the lockdown. There is also the additional dimension of studying potential differences in energy consumption patterns before, during and after lockdown periods, if any. Indoor noise levels before, during and after lockdown periods present another interesting scenario, especially in terms of correlating them to activity inside schools and seeing whether there are tangible differences that could be traced to behavior change patterns.

In addition, variations in humidity, temperature and luminosity, can be used to indirectly check for behavior changes, as well as for studying the effect of the lockdown and COVID-19 measures on indoor comfort levels inside the schools. Specifically for comfort levels, we will utilize the Predicted Mean Vote [15] (PMV) indicator to study this aspect.

The sensors in the utilized infrastructure were sampled at least every 30 seconds, with each batch of samples from all the available sensors reported back in near real time. The collected data were then aggregated and stored in our backend, either as raw data or as aggregated historical data in multiple intervals (5 minutes, 1 hour, 1 day). The analysis presented in the next section is mainly based on an extracted dataset containing the 1 hour aggregated data from all school buildings and for a period from 1/1/2019 until late May 2021. To clean and process the data in the extracted dataset we use Project Jupyter [16] in Python, together with tools like Pandas [17] and Matplotlib [18].

## V. RESULTS - DISCUSSION

### A. Power Consumption Comparison

Regarding electric power consumption in our set of school buildings, as mentioned above we utilized measurements taken every 30 seconds, which were then processed to produce daily sum values. The measurements in the majority of these school buildings represent either total building consumption, or a significant part of the building, e.g., including all but one floor of the building. In other words, we have a very representative dataset of the energy consumption in these schools, although we do not include other energy sources, such as gas. Moreover, for the examples included in our figures, which cover the period from March to May, use of gas/oil is minimal, or close to zero, in Greece in such buildings, since they are mainly used for heating.

In Fig. 3, we include some characteristic examples of the consumption of school buildings of various levels. The data displayed show average power consumption over the duration of a day, between March 23 and May 5, for 3 consecutive years. In the first 2 examples, at first glance, we see a more or less expected picture; consumption during 2020 and 2021 is significantly lesser than 2019, with the one in 2021 a bit higher compared to 2020, due to schools being open at least for some days. However, we can also spot other interesting findings, with the consumption during off-class hours being higher in 2021 than 2020, and 2019, in the first example, while in the second, 2020 displays a lower consumption overall.

Moving on to the rest of the examples, we can see a somewhat different picture. Some schools present a higher consumption during off-hours in 2020, in some cases probably due to the use of external lighting for security reasons during night time, such as in the case of school H1. In the last example, we see a school that is quite optimized for off-hours with minimal consumption, and the consumption in 2021 being close to the one of 2019, indicating quite high levels of activity.

Furthermore, when looking at the average picture of power consumption during normal periods and during the lockdown in Fig. 4, in our set of schools we see that the power consumption during night time in all 3 years is mostly similar, and is close to 40% of the typical power consumption during class time. We also see that during the lockdown of 2020 when all schools were closed down uniformly power consumption during daytime is smaller than the night time, but is again close to 30% of the normal daytime power consumption. These two findings could be explained by the use of additional lighting for security reasons during night time, and the non-optimal use of infrastructure inside the school building.

*B. Noise levels comparison*

Moving on to noise levels, several different types of hardware are utilized in the school infrastructure, due to the fact that the deployment roll-out spanned across a number of years. Such hardware ranges from very basic types of audio sensors to purpose-built noise level sensors. In this sense, it is difficult to draw comparisons across the whole set of schools, however it should be sufficient for comparing measurements in a relative manner, spanning across the 3 last years for more general observations.

Fig. 5 and Fig. 6 provide an overview of the noise levels in several schools between January 2019 and May 2020 and during the 2020-21 respectively. We can observe that noise levels align almost perfectly with the activity in schools, break periods and the lockdown. In this sense, we can use the noise levels to further characterize the activity inside school buildings, and get a better sense of the overall activity levels in the schools, together with the power consumption data. Another interesting finding is that here are examples of noise levels being both similar and lower after the lockdown period. Lower noise levels could also be attributed to certain classes being shut down due to COVID-19 cases. In Fig. 6 such differences are more pronounced, with some schools being completely shut down, while others had some range of activity. We can also some bigger noise level spikes in 2 schools after reopening, while in the other noise spikes are similar or a bit smaller than the pre-lockdown levels. Therefore, at this point, results seem to be on a school-by-school basis.

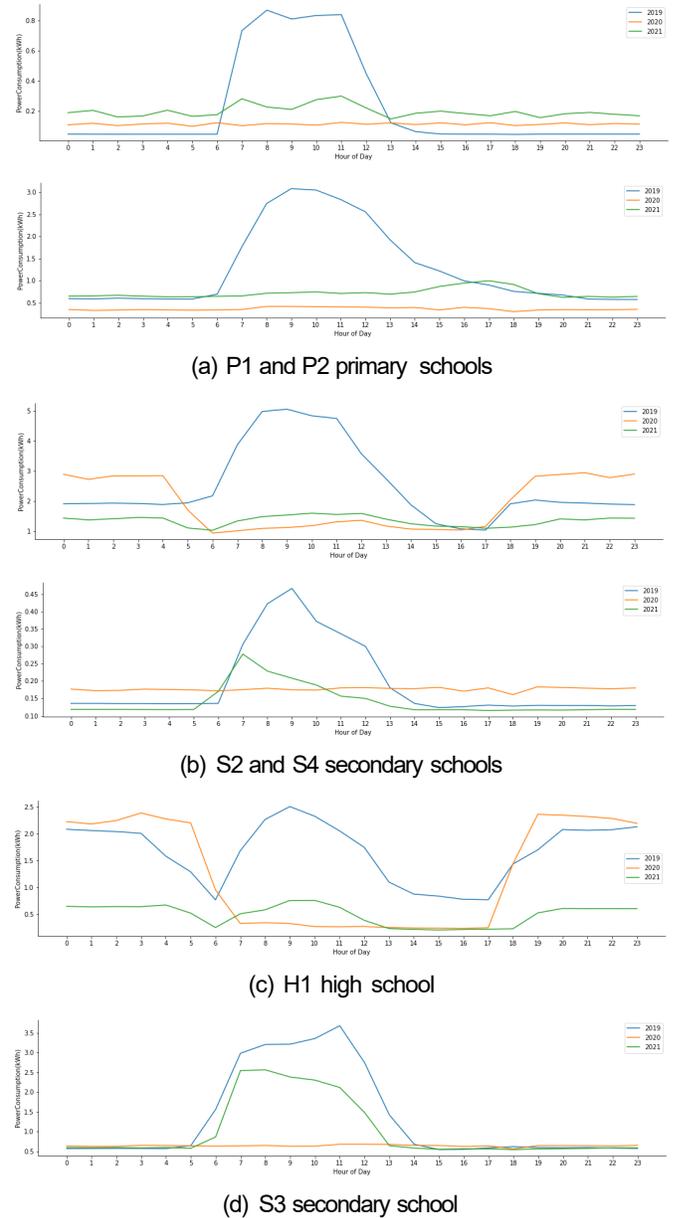

(a) P1 and P2 primary schools

(b) S2 and S4 secondary schools

(c) H1 high school

(d) S3 secondary school

Fig. 3: Average Hourly Power Consumption for schools between 23/3 and 5/5 for 2019, 2020 and 2021.

*C. Indoor comfort levels and PMV*

PMV indicates comfort for a group of subjects given a particular combination of air temperature, mean radiant temperature, relative humidity, air speed, metabolic rate, and clothing. The PMV implementation we use for this analysis uses an eleven-point scale from cold (−5) to hot (+5) to represent the thermal conditions inside school buildings with more detail in contrast to other implementations with that use the [-3,3] range. A PMV equal to 0 represents thermal neutrality, and the comfort zone is defined by the combinations of the 6 parameters for which PMV is between 1 and +1.

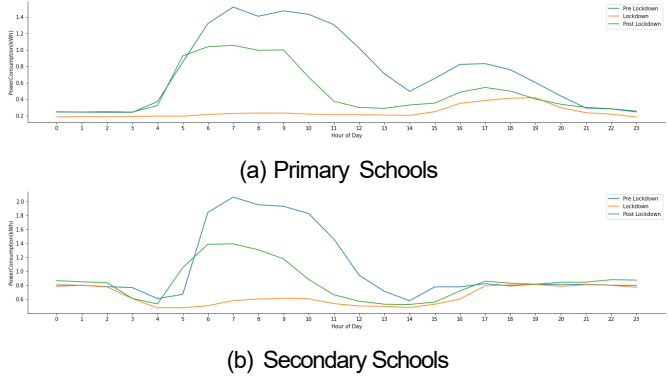

(a) Primary Schools

(b) Secondary Schools

Fig. 4: Average Hourly Power Consumption for primary and secondary schools before, during and after the 2020 42-day lockdown.

Observing the data presented in Fig. 7, we can observe how during the first lockdown period in 2020, the conditions in all schools deteriorated, as expected, as the schools were completely closed and heating was switched off completely. As this period was during the end of the winter, the PMV value decreased from an average around -1 to a value lower than -3 in most cases. After that negative peak, conditions gradually improved as the weather conditions were better over time.

For the second lockdown, starting in November 2020, the PMV value started again from a value around -1 to a slightly lower value. This time around the conditions did not deteriorate so fast, as conditions in Greece were still good, with not so low environmental temperatures. Once the primary schools re-opened in January 2021 (middle of winter for Greece), the schools were ordered to keep their windows open at all times to help ventilation. This lead to significant variations in the conditions inside the school buildings as heating was used to keep the classrooms hot enough for the students while the windows were open for more time and heat losses were expected. This behavior is more or less observed mostly in primary schools. Secondary schools have more consistent PMV values for the duration of the lockdown, as the buildings were closed for the whole period.

## VI. CONCLUSIONS

In this work, we focused on studying the effect of the COVID-19 pandemic on a diverse set of schools in Greece. This was made possible via an IoT infrastructure installed in previous years for the purposes of sustainability awareness and energy savings, which provided us with datasets containing power consumption, noise and air quality levels, temperature, humidity and luminosity data. From our results, it is obvious that the way school buildings operate could be further optimized in terms of energy efficiency. During the period of the lockdown, buildings consumed a significant amount of energy close to, or exceeding, 20-40% of the energy consumed on average in normal periods. Furthermore, regarding noise levels they can be used to infer the levels of activity inside the schools, together with power consumption, but results are inconclusive at this point in our study as regards behavior change. Moreover, indoor conditions, as regards comfort, have shifted considerable during the lockdown. Regarding our future work, we intend to delve deeper into the available datasets and further study the impact of the pandemic on school operation, and especially as regards potential behavior changes in students and educators.

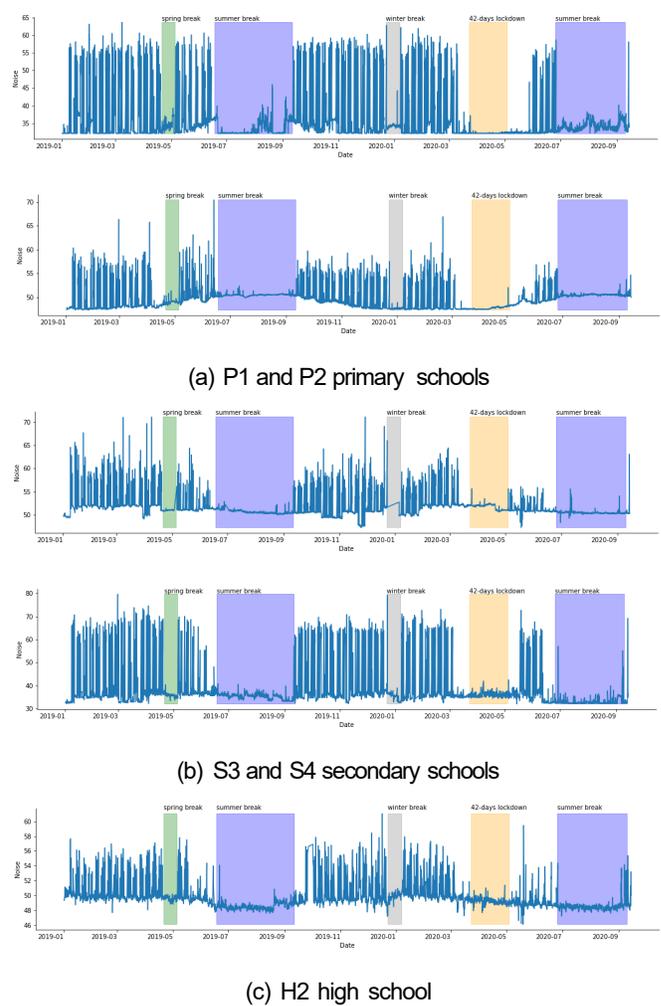

(a) P1 and P2 primary schools

(b) S3 and S4 secondary schools

(c) H2 high school

Fig. 5: Noise levels during 2019 and 2020 lockdown on 5 schools of all grades.

### ACKNOWLEDGMENT

This work has been supported by the European Union's research project "European Extreme Performing Big Data Stacks" (E2Data), funded by the European Commission (EC) under the Horizon 2020 framework and contract number 780245, and the "Green Awareness In Action" (GAIA) project, funded by the European Commission and the Executive Agency for Small and Medium-sized Enterprises (EASME) under the Horizon 2020 framework and contract number 696029. This document reflects only the authors' views and the EC and EASME are not responsible for any use that may be made of the information it contains.

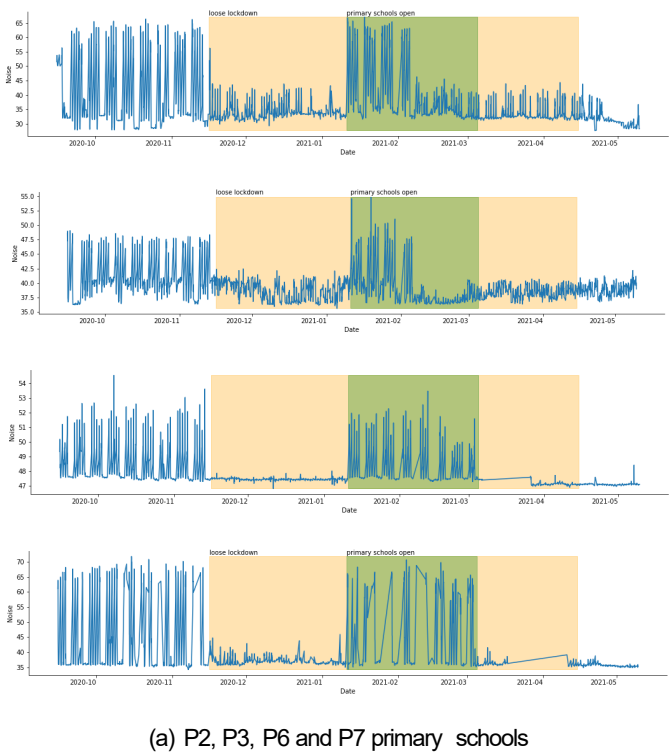

(a) P2, P3, P6 and P7 primary schools

Fig. 6: Noise levels during the 2020-21 lockdown for 4 different primary schools.

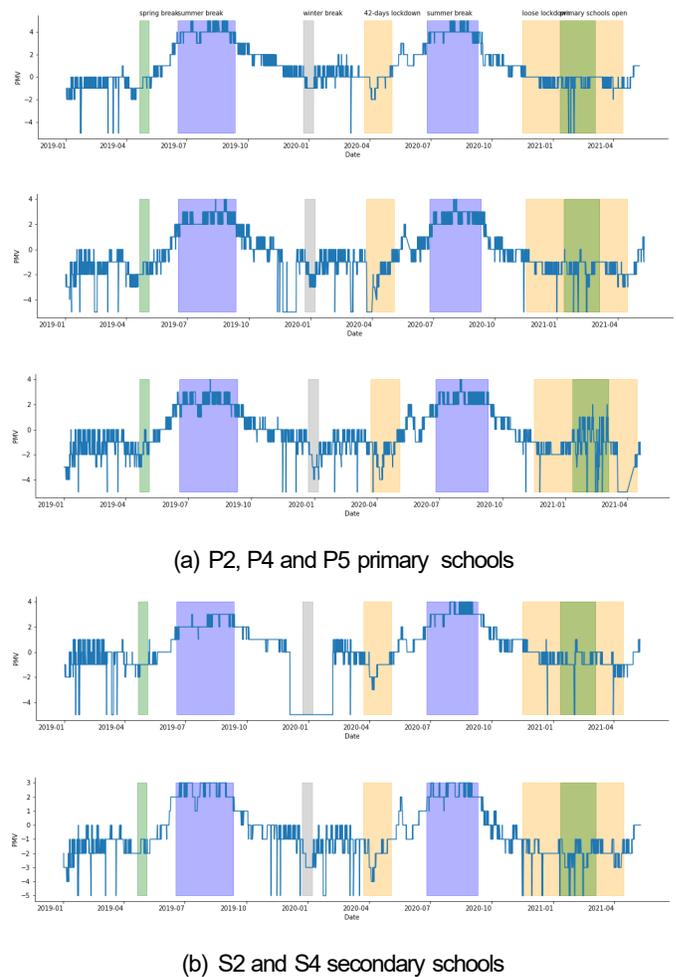

(a) P2, P4 and P5 primary schools

(b) S2 and S4 secondary schools

Fig. 7: PMV values between January 2019 and May 2021.